\documentclass[cameraready]{Interspeech}

\title{Task-Aware Joint Pruning and Distillation for Efficient Audio Deepfake Detection}

\author[affiliation={1}]{Miao}{He}
\author[affiliation={1,2}, correspondingauthor]{Peng}{Cheng}
\author[affiliation={1,2}]{Zhongjie}{Ba}
\author[affiliation={1}]{Qing}{Wen}
\author[affiliation={1,2}]{Li}{Lu}
\author[affiliation={4}]{Xin}{Yang}
\author[affiliation={1,2,3}]{Kui}{Ren}

\address{
    $^1$ The State Key Laboratory of Blockchain and Data Security, Zhejiang University, China \\
    $^2$ Hangzhou High-Tech Zone (Binjiang) Institute of Blockchain and Data Security, China \\
    $^3$ Shanghai Institute for Advanced Study, Zhejiang University, China \\
    $^4$ Qingdao Institute of Software College of Computer Science and Technology, China University of Petroleum (East China), China
}

\email{miao\_he@zju.edu.cn, peng\_cheng@zju.edu.cn, zhongjieba@zju.edu.cn}


\usepackage{comment}

\usepackage{booktabs}
\usepackage{multirow}
\usepackage{makecell}
\usepackage{graphicx}
\usepackage{array}
\usepackage{subcaption} 
\usepackage{colortbl}

\begin{document}

\maketitle

\begin{abstract}

     Advances in speech synthesis have made deepfake speeches increasingly convincing, posing growing threats to security. While self-supervised learning (SSL) based detectors achieve state-of-the-art performance, their computational demands (typically 300M+ parameters) prevent deployment on resource-constrained devices. Existing compression methods, designed mainly for content-centric tasks, struggle to maintain competitive performance when directly adapted to deepfake detection. We propose a Task-Aware Joint Pruning and Distillation framework that combines cross-domain knowledge distillation with movement-guided structured pruning to transfer forgery-discriminative knowledge and preserve critical structures under aggressive compression. Our framework reduces the model to 31.9M parameters with 6.3$\times$ FLOPs reduction, with an average performance drop of only 1.30\% across multiple datasets compared to the uncompressed baseline, demonstrating strong potential for on-device deployment.
  
\end{abstract}

\section{Introduction}

With the rapid evolution of speech synthesis technology, generative models can now synthesize high-fidelity speech that is perceptually indistinguishable from bonafide audio~\cite{wang2025one, warren2024better}, posing an urgent need to develop reliable countermeasures. The synthesized speeches, namely deepfake speeches, can deceive both human and Automatic Speaker Verification systems, posing a severe threat to security, with real-world incidents already demonstrating malicious exploitation of voice cloning for telecommunications fraud and identity theft~\cite{2024dangers,npr2025Eddie_deepfake_advertisement,npr2025impostor_ai_rubio}. To mitigate these threats, self-supervised learning (SSL) models—such as wav2vec2.0~\cite{baevski2020wav2vec}, XLSR~\cite{babu2021xls}, and WavLM~\cite{chen2022wavlm}—have emerged as the dominant feature extractor for deepfake detection, achieving state-of-the-art performance~\cite{tak2022automatic, zhang2024audio, rosello2023conformer, xiao2025xlsr-mamba, dowerah2026speechBoard}.

However, the immense computational complexity of SSL-based detectors (typically 300M+ parameters) precludes their deployment on resource-constrained devices, necessitating cloud-based inference. This introduces two critical bottlenecks: (1) \textit{Network Dependency}: cloud inference requires stable connectivity, compromising availability in unstable network environments; (2) \textit{Privacy Risks}: uploading user audio to cloud servers raises privacy concerns~\cite{wang2011privacy, yu2015enabling}. A promising solution to address these bottlenecks is on-device deepfake detection, which requires lightweight models that can operate on resource-limited devices while maintaining competitive performance.


Recent efforts have investigated compressing speech SSL models through knowledge distillation~\cite{chang2022distilhubert, wang2023distilxlsr, lee2022fithubert} and structured pruning~\cite{peng2023structured, Han2025FinetuneBS, peng2025hybrid, peng2023dphubert}. However, directly adapting these methods to deepfake detection faces critical challenges. First, existing compression strategies are predominantly tailored for content-centric tasks like ASR, leaving deepfake-specific compression largely unexplored. Second, existing techniques rely solely on task-specific loss or in-domain distillation as optimization objectives. When directly adapted to deepfake detection, such limited supervision fails to transfer sufficient discriminative knowledge and preserve critical structures, leading to poor generalization. Third, we also empirically observe that current methods degrade severely at high compression ratios ($>$75\% sparsity), especially in out-of-domain scenarios. 

To address these limitations, we propose a \textbf{Task-Aware Joint Pruning and Distillation} framework specifically designed for compressing SSL-based deepfake detection models. We introduce \textit{cross-domain knowledge distillation} to transfer forgery-discriminative representations and enhance generalization by exploiting diverse unlabeled out-of-domain data as richer supervisions, with representative layers selected via CKA analysis for multi-level knowledge distillation. We further adopt \textit{movement-guided structured pruning} that leverages fine-tuning movement score to prioritize the preservation of forgery-critical structures during compression. 
Extensive experiments demonstrate that our framework maintains competitive performance even under aggressive compression, with an average performance drop of only 1.30\% across multiple datasets compared to the uncompressed baseline. Crucially, our compressed model achieves a 10$\times$ parameter reduction (31.9M) and 6.3$\times$ FLOPs reduction compared to the uncompressed baseline, demonstrating strong potential for on-device deployment.

\section{Background and Related Work}

\noindent\textbf{Speech SSL for Deepfake Detection.}
Self-supervised learning models have been extensively explored for audio deepfake detection in recent years \cite{tak2022automatic, zhang2024audio, rosello2023conformer, xiao2025xlsr-mamba}. These SSL frontends, typically comprising CNNs and Transformers, leverage large-scale pretraining to learn rich representations, making them effective for deepfake detection \cite{dowerah2026speechBoard}. However, their computational demands (typically 300M+ parameters, 130+ GFLOPs) prevent deployment on resource-constrained devices. 


\noindent\textbf{Compression Methods for Speech SSL.}
Speech SSL model compression primarily relies on knowledge distillation and structured pruning. \textit{Knowledge distillation} transfers knowledge to student models via layer-wise distillation~\cite{chang2022distilhubert, wang2023distilxlsr, lee2022fithubert}, yet requires manual architecture design. While recent works~\cite{xue2023learning,lu2024one} have applied distillation to deepfake detection, they optimize merely for accuracy without compressing for on-device deployment. Conversely, \textit{structured pruning} automatically removes redundant units through task-agnostic~\cite{peng2023dphubert, wang2023task} or task-specific approaches~\cite{peng2023structured, Han2025FinetuneBS, peng2025hybrid}. However, adapting these methods to deepfake detection leads to severe performance degradation at high compression ratios. Our proposed framework is specifically designed to address these limitations.

\begin{figure}[t]
  \centering
  \includegraphics[width=\linewidth]{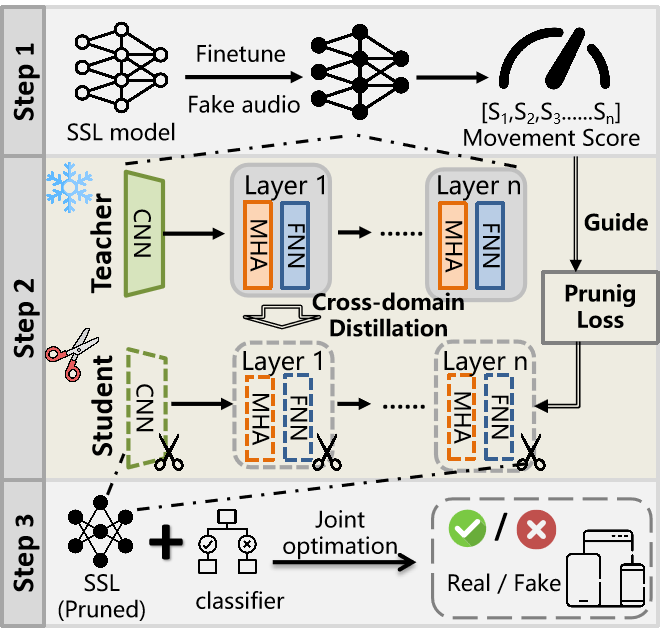}
  \caption{Task-Aware Joint Pruning and Distillation framework: (1) Task-specific preparation; (2) Joint pruning and distillation; (3) Backend optimization. The compressed model enables efficient on-device deepfake detection.}
  \label{fig:method}
  \vspace{-0.5cm} 
\end{figure}

\begin{table*}[t]
\centering
\caption{Performance comparison between our method and baselines across various sparsity levels (60\%, 75\%, and 90\%). EER (\%) values are reported, with $\downarrow$ indicating lower is better. \textbf{Bold} and \underline{underline} indicate the best and second-best results, respectively}
\label{tab:EER_results}
\setlength{\tabcolsep}{4.5pt} 
\footnotesize 
\begin{tabular}{clcccccccccc}
\toprule
\multirow{2}{*}{\textbf{Sparsity}} & \multirow{2}{*}{\textbf{Methods}} & \multirow{2}{*}{\textbf{\#Param}} & \multirow{2}{*}{\textbf{FLOPs}} & \multicolumn{7}{c}{\textbf{Evaluation Datasets (EER \%) $\downarrow$}} \\
\cmidrule(lr){5-11}
& & & & \textbf{19LA\_Dev} & \textbf{19LA\_Eval} & \textbf{21LA} & \textbf{21DF} & \textbf{In-the-Wild} & \textbf{ASV5} & \textbf{FoR} \\
\midrule
\rowcolor{gray!20}
0\% & XLSR-AASIST & 317 M & 146.3 G & 0.04 & 0.14 & 2.86 & 3.09 & 8.96 & 19.92 & 14.89 \\
\midrule
\multirow{4}{*}{60\%}  
& HJ-Pruning\cite{peng2023structured} & 127.9 M & 67.9 G & \underline{0.04} & 2.40 & 14.20 & 21.27 & 33.61 & 38.79 & 59.49 \\

& Finetune-Pruning\cite{Han2025FinetuneBS} & 126.3 G & 63.5 G & 0.08 & \underline{0.20} & \underline{7.64} & \underline{3.99} & \underline{17.22} & \underline{24.10} & \underline{16.52} \\
& Hybrid-Pruning\cite{peng2025hybrid} & 125.6 M & 69.5 G & \underline{0.04}  & 17.93 & 25.98 & 30.61 & 53.27 & 45.17 & 62.23 \\
& \textbf{Ours} & 126.6 M & 63.8 G & \textbf{0.03} & \textbf{0.18} & \textbf{3.24} & \textbf{2.21} & \textbf{7.29} & \textbf{16.46} & \textbf{9.36} \\
\midrule
\multirow{4}{*}{75\%}
& HJ-Pruning &79.4 M&46.6 G& \underline{0.03} & 2.73 & 13.53 & 22.37 & 29.66 & 38.44 & 56.58 \\
& Finetune-Pruning & 79.0 M & 43.7 G & 0.04 & \underline{0.38} & \underline{9.56} & \underline{8.23} & \underline{25.50} & \underline{26.13} & \underline{14.80} \\
& Hybrid-Pruning & 78.1 M & 50.3 G & 0.04 & 13.19 & 25.59 & 30.50 & 39.67 & 44.73 & 63.78 \\
& \textbf{Ours} & 79.2 M & 43.9 G & \textbf{0.004} & \textbf{0.22} & \textbf{3.03} & \textbf{2.76} & \textbf{8.43} & \textbf{17.75} & \textbf{10.51} \\
\midrule
\multirow{4}{*}{90\%}
& HJ-Pruning &32.6 M& 27.1 G & 0.08 & 5.03 & 15.88 & 23.30 & 34.11 & 40.82 & 57.55 \\
& Finetune-Pruning & 31.3 M & 22.2 G & \underline{0.04} & \underline{0.52} & \underline{10.09} & \underline{12.06} & \underline{32.54} & \underline{35.20} & \underline{27.43} \\
& Hybrid-Pruning & 29.2 M & 29.5 G & 0.12 & 10.39 & 24.55 & 26.94 & 52.06 & 43.78 & 58.61 \\
& \textbf{Ours} & 31.9 M & 23.3 G & \textbf{0.002} & \textbf{0.30} & \textbf{3.11} & \textbf{3.78} & \textbf{7.83} & \textbf{20.16} & \textbf{22.44} \\
\bottomrule
\end{tabular}
\vspace{-0.2cm}
\end{table*}

\begin{figure*}[t!] 
  \centering
  
  \begin{subfigure}[b]{0.32\linewidth}
    \centering
    \includegraphics[width=\linewidth]{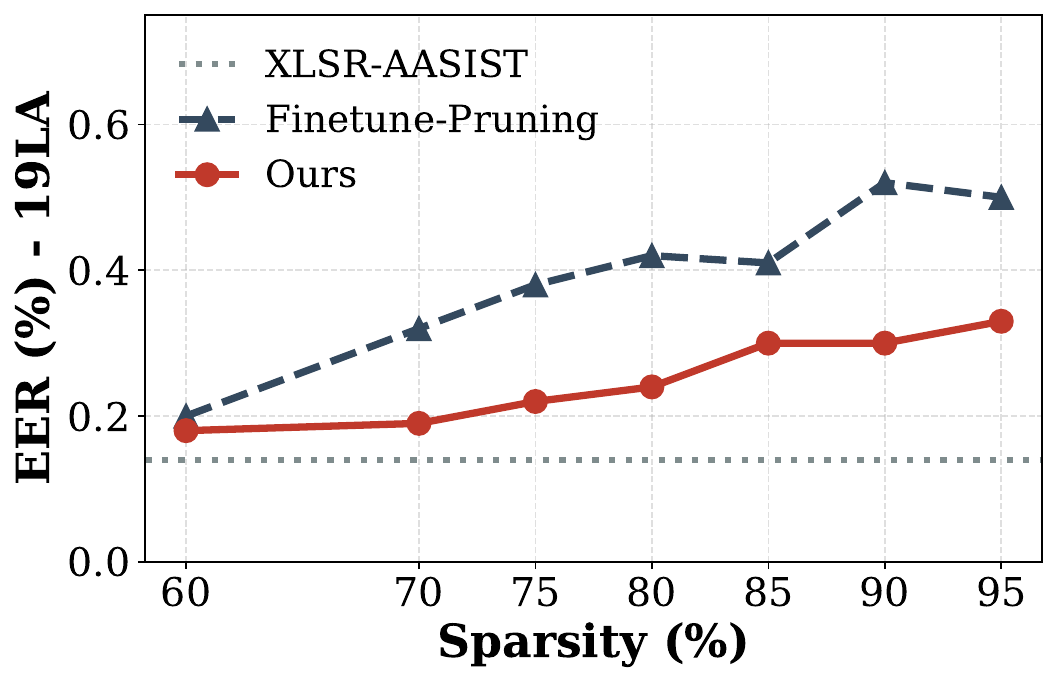} 
    \caption{ASVspoof2019 LA Eval}
    \label{fig:i19LA}
  \end{subfigure}
  \hfill 
  \begin{subfigure}[b]{0.32\linewidth}
    \centering
    \includegraphics[width=\linewidth]{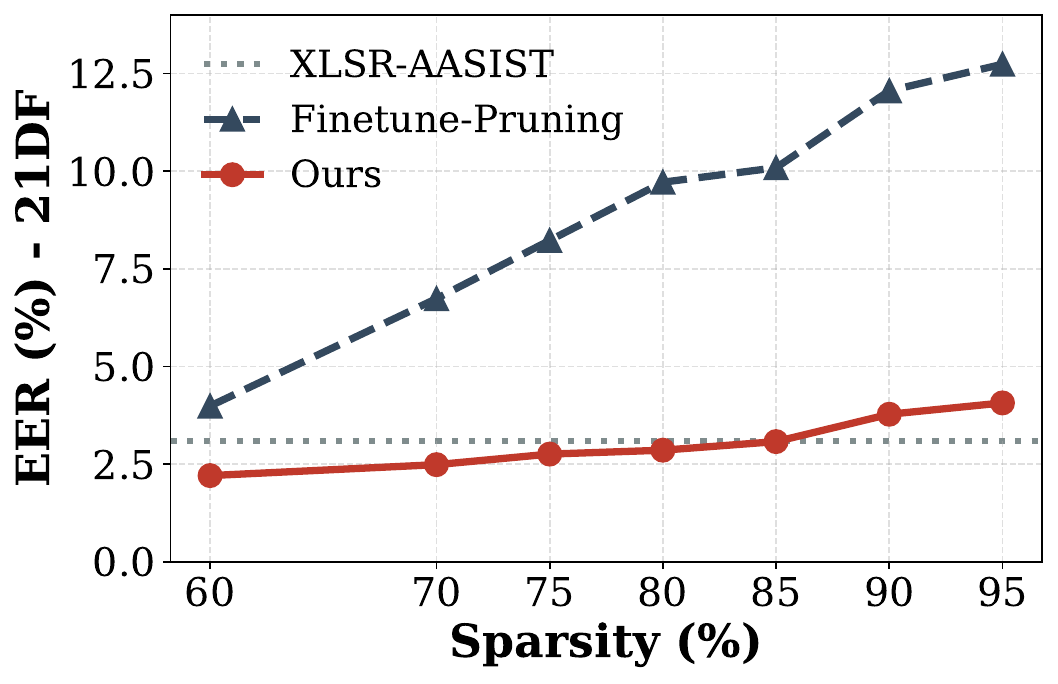}
    \caption{ASVspoof2021 DF}
    \label{fig:21DF}
  \end{subfigure}
  \hfill 
  \begin{subfigure}[b]{0.32\linewidth}
    \centering
    \includegraphics[width=\linewidth]{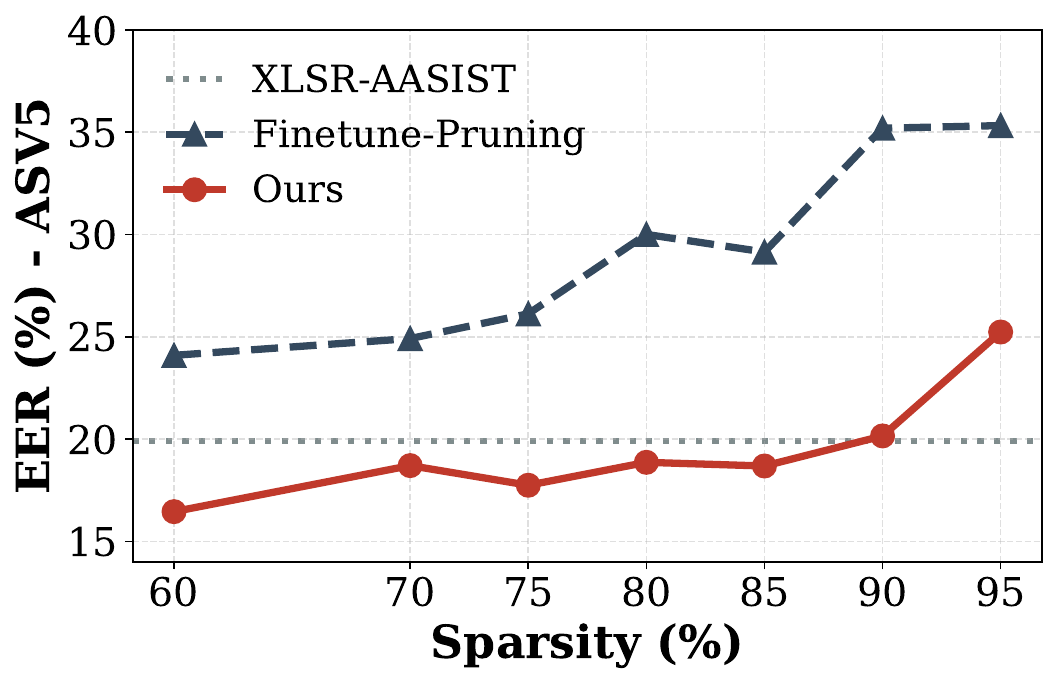}
    \caption{ASVspoof5}
    \label{fig:ASVspoof5}
  \end{subfigure}
  \vspace{-0.2cm} 
  
  \caption{Model performance (EER\%) changes as sparsity increases on (a) ASVspoof2019 LA Eval, (b) ASVspoof2021 DF, and (c) ASVspoof5. Our method (red line) consistently outperforms the Finetune-Pruning method (blue dashed).}
  
  \label{fig:curves}
  \vspace{-0.5cm}
\end{figure*}

\begin{table}[th]
  \centering
  \caption{Ablation study at 75\% sparsity. Results in EER (\%).}
  \label{tab:ablation}
  
  \footnotesize
  \setlength{\tabcolsep}{3.5pt} 

  \begin{tabular*}{\linewidth}{lcccccc}
    \toprule
    \textbf{Config.} & \textbf{19 Eval} & \textbf{21LA} & \textbf{21DF} & \textbf{ITW} & \textbf{ASV5} & \textbf{FoR} \\
    
    \midrule
    \rowcolor{gray!20}
    XLSR-AASIST & 0.14& 2.86 & 3.09 & 8.96 & 19.92 & 14.89 \\
    \midrule
    
    \textbf{Ours (Full)} & \textbf{0.22} & \underline{3.03} & \textbf{2.76} & \underline{8.43} & \textbf{17.75} & \textbf{10.51} \\
    
    \midrule
    \ w/o Cross. Distill & 0.39 & 10.14 & 8.25 & 25.84 & 27.48 & 16.61 \\
    \ w/o Move. Guide & \underline{0.30} & \textbf{3.00} & \underline{3.13} & \textbf{7.83} & \underline{19.95} & \underline{13.56} \\ 
    \bottomrule
  \end{tabular*}
  \vspace{-0.3cm}
\end{table}

\section{Methodology}
As illustrated in Figure \ref{fig:method}, our proposed framework consists of three steps: (1) Task-specific preparation: fine-tune the pretrained SSL model and compute movement scores of structural units as an importance prior, including Multi-Head Attention, FFN intermediates, and CNN channels; (2) Joint pruning and distillation: the structured pruning is guided by the movement importance prior, combined with cross-domain knowledge distillation to preserve forgery-discriminative structures; (3) Backend optimization: jointly fine-tune the pruned SSL with the backend classifier to optimize detection accuracy.

\subsection{Cross-Domain Knowledge Distillation}
Unlike existing task-specific structured pruning with distillation methods that distill on in-domain training data \cite{peng2023structured,Han2025FinetuneBS}, we adopt a cross-domain distillation strategy to maintain both accuracy and generalization under compression.


Specifically, we leverage \textit{unlabeled out-of-domain task-relevant} data covering diverse codec variations and environmental distortions to provide richer supervisions during compression. Since our uncompressed teacher model already exhibits reasonable generalization on several out-of-domain scenarios (EER\textless 10\%), enforcing the student to mimic the teacher's representations across these diverse scenarios encourages learning of generalizable forgery-discriminative representations rather than overfitting to training distribution patterns. Specifically, we incorporate ASVspoof2021 LA, 21 DF and In-the-Wild datasets as the out-of-domain distillation datasets.

To further improve distillation efficiency, we analyze layer similarity via Centered Kernel Alignment (CKA) \cite{pmlr-v97-kornblith19a}. As shown in Figure~\ref{fig:cka_comparison}, CKA analysis of the finetuned XLSR reveals three distinct block-like structures corresponding to different processing stages: Layers 1--8, 9--21, and 22--24.  Guided by prior findings that lower layers--particularly the 5th layer--capture discriminative features \cite{el2025comprehensive,lee2022representation}, we select representative layers in each block (e.g., $\mathcal{L}_{set}=\{5, 14, 24\}$) for multi-level knowledge transfer:
\begin{align}
  \mathcal{L}_{distill} = \sum_{l \in \mathcal{L}_{set}}L^{dis}(f_{T}^l(x), \phi(f_{S}^l(x)))
  \label{equation:eq2}
\end{align}
where $f_{T}^l$ and $f_{S}^l$ denote teacher and student features at layer $l$, $\phi(\cdot)$ is a dimension alignment projection, and $L^{dis}$ is MSE or cosine distance. Here $x$ spans both in-domain and cross-domain samples without labels, enabling the pruned model to preserve generalization even at high compression ratios.

\subsection{Movement-Guided Structured Pruning}
Beyond cross-domain distillation, we further introduce movement-guided pruning to better preserve task-aware structures. Previous work has shown that weights exhibiting significant changes (i.e., movement) during fine-tuning encode task-specific knowledge~\cite{sanh2020movement}. Building on this insight, we define a Movement Score $S_j = ||\theta_j^{ft} - \theta_j^{init}||_1$ for each structural unit $j$ and assign an inverse penalty weight $w_j$, yielding an importance-aware regularization term:
\begin{align}
  \mathcal{L}_{penalty} = \sum_{j=1}^n w_j \cdot z_j, \quad \text{where } w_j = \frac{1}{S_j + \epsilon}
  \label{equation:eq1}
\end{align}
where $z_j \in \{0,1\}$ is the binary retention mask. Minimizing $\mathcal{L}_{penalty}$ encourages the model to prune low-movement redundant units while retaining those essential for deepfake detection.

\subsection{Joint Optimization Objective}
We formulate the training as a constrained multi-objective optimization to achieve the target compression ratio $t$ as follows:
\begin{align}
  \min_{\theta, \mathbf{z}} \left[ \mathcal{L}_{distill} + \lambda_{p} \mathcal{L}_{penalty} \right] \quad 
  \text{s.t.} \quad \frac{||\theta_j \odot \mathbf{z}_j||_0}{N} = t
  \label{equation:eq3}
\end{align}
Following \cite{peng2023dphubert, louizos2018learning}, we reparameterize the non-differentiable mask $\mathbf{z}_j$ via the Hard Concrete distribution, making the $L_0$ norm differentiable w.r.t. learnable parameters $\alpha$:
\begin{align}
    \mathbb{E}[||\tilde{\theta}||_0] = \sum_{j=1}^n \sigma(\log \alpha_j - \beta \log \frac{-l}{r})
\end{align}
where $\tilde{\theta}=\{\theta_j \odot \mathbf{z}_j\}_{j=1}^n$, and $\beta$ is a temperature constant, and $l{<}0$, $r{>}0$ stretch the sigmoid to $[l,r]$, rectified to $[0,1]$.
The current sparsity is then $s(\alpha) = \mathbb{E}[||\tilde{\theta}||_0]\ /\ N$. 
To strictly enforce target sparsity $t$, we apply the Augmented Lagrangian method \cite{peng2023dphubert,wang2020structured,xia2022structured} by introducing Lagrangian multipliers $(\lambda_1, \lambda_2)$, from which we derive the final optimization objective:
\begin{align}
  \max_{\lambda_1,\lambda_2} \min_{\theta,\alpha} \Bigg[
  \underbrace{\mathcal{L}_{distill}}_{\text{Distillation}} + 
  \underbrace{\lambda_p \mathcal{L}_{penalty}}_{\text{Task-Aware Prior}} + 
  \underbrace{\mathcal{L}_{constraint}}_{\text{Target Sparsity}}
  \Bigg] \nonumber \\
  \text{where  } \mathcal{L}_{constraint} = \lambda_1(s(\alpha) - t) + \lambda_2(s(\alpha) - t)^2
  \label{equation:eq6}
\end{align}


\section{Experiments}
\subsection{Experimental Setup}
\noindent\textbf{Datasets and Evaluation.} 
We use the ASVspoof2019 LA~\cite{todisco2019asvspoof} training set for fine-tuning in step 1 and step 3. For distillation, we adopt the 19LA train, ASVspoof2021 LA, 21 DF~\cite{yamagishi2021asvspoof}, and In-the-Wild~\cite{muller2022does-wild} without labels. We evaluate on in-domain 19LA Dev and Eval; cross-domain distillation benchmarks: ASVspoof2021 LA, 21DF, and In-the-Wild; and totally unseen benchmarks: ASVspoof5~\cite{wang2024asvspoof5} and FoR~\cite{reimao2019dataset-for}. We report Equal Error Rate (EER) for detection performance, model size (Params) and computation cost (FLOPs) for efficiency. 

\noindent\textbf{Implementation Details.} 
We compress the XLS-R (0.3B)~\cite{babu2021xls} of the XLSR-AASIST~\cite{tak2022automatic} backbone. 
For pruning and distillation, the student is trained for 50k steps, with target sparsity linearly increasing to $t$ during a 5k warmup. Learning rates are set to 2e-4 for model parameters $(\theta)$ and 2e-2 for auxiliary parameters $(\alpha, \lambda_1, \lambda_2)$. Following~\cite{peng2023dphubert}, the pruned student is further distilled for 25k steps, and finally jointly fine-tuned with the AASIST backend for a few epochs.

\noindent\textbf{Baselines.} 
We compare against the original XLSR-AASIST baseline and three pruning methods originally designed for other speech tasks: 
(1) \textbf{HJ-Pruning}~\cite{peng2023structured}: Heterogeneous joint pruning of CNN and Transformer layers for speech recognition; 
(2) \textbf{Finetune-Pruning}~\cite{Han2025FinetuneBS}: Fine-tune before structured pruning for speaker diarization;
(3) \textbf{Hybrid-Pruning}~\cite{peng2025hybrid}: Hybrid in-situ compression for speaker verification and anti-spoofing. 
All methods are re-implemented for the deepfake detection task under a unified framework for fair comparison.

\subsection{Main Results}
As shown in Table~\ref{tab:EER_results}, our method consistently outperforms all baselines across varying sparsity levels, particularly on out-of-domain datasets. At 75\% sparsity, for instance, our method achieves 8.43\% EER on In-the-Wild and 17.75\% on ASVspoof5, outperforming the second-best baseline \textit{Finetune-Pruning} by 67\% and 32\% respectively. This demonstrates that our framework effectively maintain both accuracy and generalization.
Remarkably, our pruned models even outperform the uncompressed baseline on several out-of-domain benchmarks. For instance, at 75\% sparsity, our method achieves 17.75\% EER on ASVspoof5 and 10.51\% on FoR, surpassing the dense model's 19.92\% and 14.89\%, respectively. Similar to observations in~\cite{peng2023structured, peng2025hybrid}, this counterintuitive result suggests that the over-parameterized baseline overfits to the source domain. Pruning acts as a structural regularization, removing redundant capacity and forcing the model to retain discriminative forgery artifacts during compression.

The robustness of our approach becomes evident at even higher compression rates. At 90\% sparsity, our framework retains competitive performance with an average EER increase of only 1.30\% across multiple datasets, while other methods suffer catastrophic degradation. The pruned model achieves 10$\times$ parameters and 6.3$\times$ FLOPs reduction (31.9M parameters, 23.3G FLOPs) over the 317M/146.3G baseline, making it highly promising for resource-constrained deployment.



Figure~\ref{fig:curves} visualizes how model performance changes as sparsity increases, comparing our method against the second-best \textit{Finetune-Pruning} across three representative scenarios: ASVspoof2019 LA Eval as the in-domain set, ASVspoof2021 DF as the cross-domain distillation set, and ASVspoof5 as a totally unseen evaluation scenario. On the in-domain 19LA (Figure~\ref{fig:curves}(a)), both methods maintain strong performance with EER\textless0.6\%. However, the contrast becomes stark on out-of-domain datasets (Figure~\ref{fig:curves}(b-c)). \textit{Finetune-Pruning} degrades sharply beyond 70\% sparsity, whereas our method remains competitive with the uncompressed baseline even at sparsity beyond 90\%. These results further validate that our framework effectively addresses the generalization bottleneck of existing compression methods under high compression ratios.

\begin{figure}[t]
  \centering
  \begin{subfigure}[b]{0.48\linewidth}
    \centering
    \includegraphics[width=\linewidth]{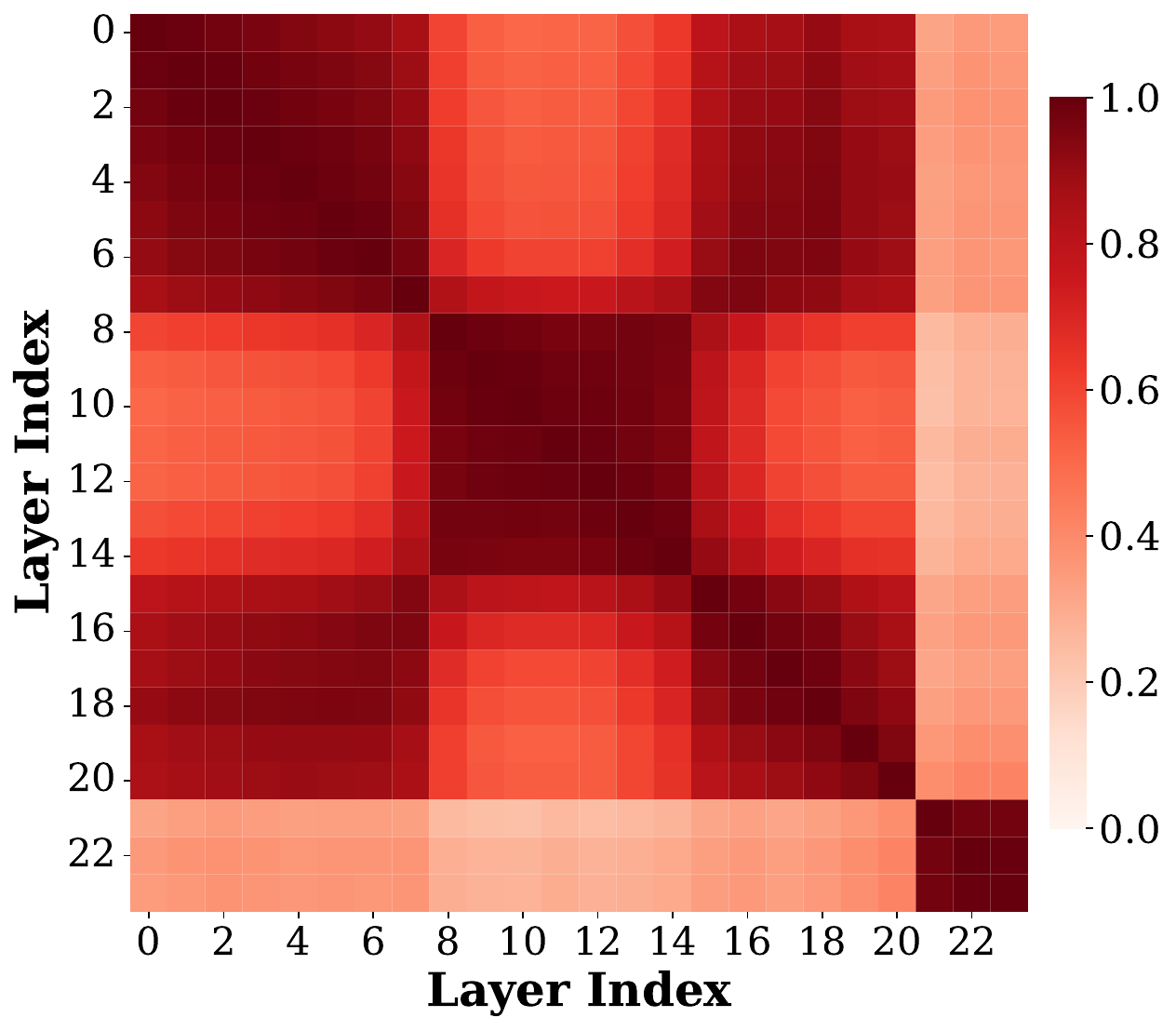} 
        \caption{Pre-trained XLSR} 
    \label{fig:cka_pre}
  \end{subfigure}
  \hfill 
  \begin{subfigure}[b]{0.48\linewidth}
    \centering
    \includegraphics[width=\linewidth]{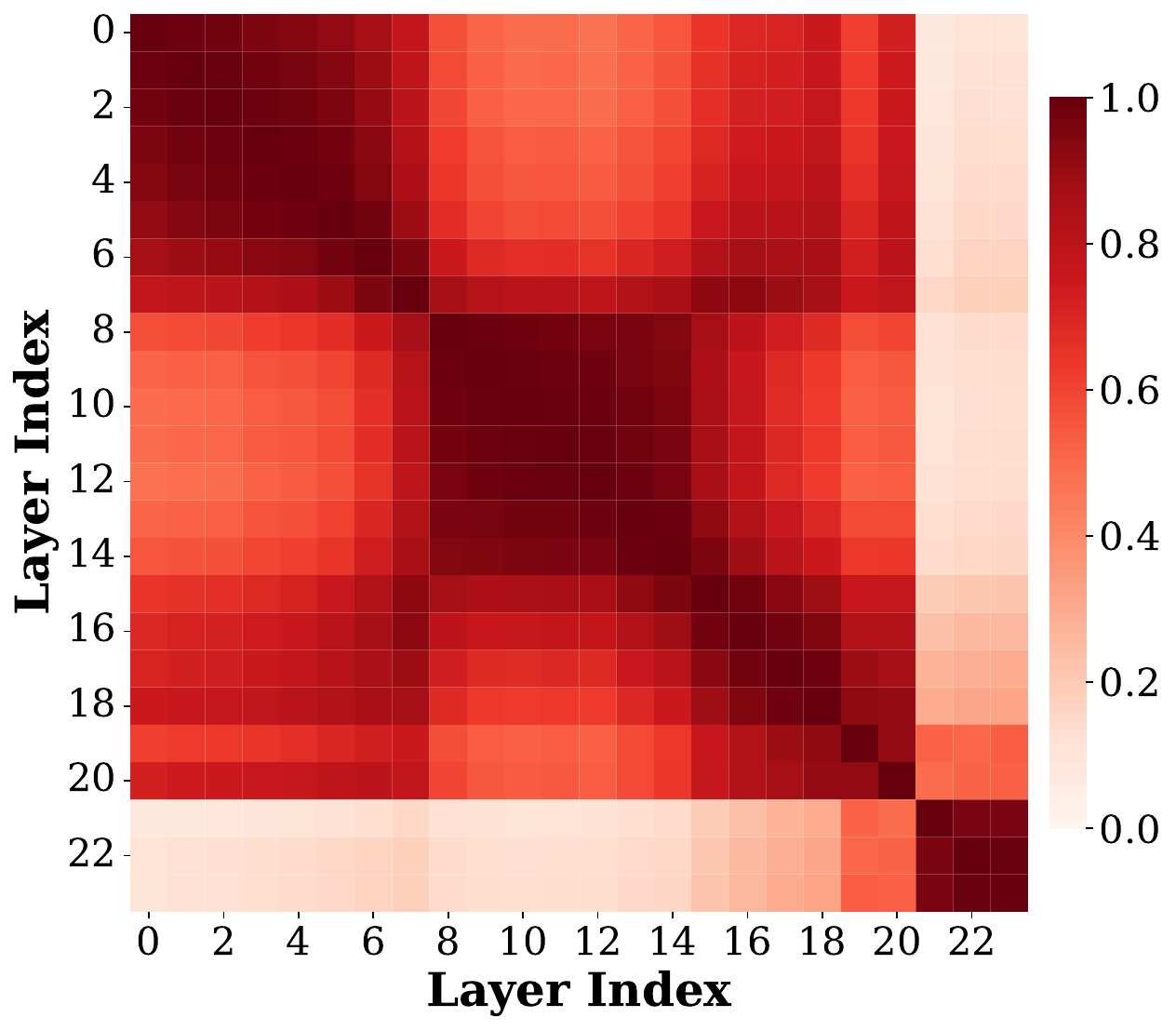}
    \caption{Finetuned XLSR} 
    \label{fig:cka_finetune}
  \end{subfigure}
  
  \vspace{-0.2cm} 
  \caption{Layer-wise CKA similarity for (a) Pre-trained XLSR and (b) Finetuned XLSR.}
  \label{fig:cka_comparison}
  \vspace{-0.7cm} 
\end{figure}

\subsection{Ablation Studies}
We conduct an ablation analysis at 75\% sparsity in Table \ref{tab:ablation} to dissect the explicit contributions of each core component.

\noindent\textbf{Cross-Domain Distillation.} 
This is the primary driver of generalization. Removing it causes severe degradation on out-of-domain datasets, such as In-the-Wild (8.43\% $\rightarrow$ 25.84\%), ASVspoof5 (17.75\% $\rightarrow$ 27.48\%), and ASVspoof2021 LA (3.03\% $\rightarrow$ 10.14\%). Meanwhile, in-domain performance on 19LA Eval degrades minimally (0.22\% $\rightarrow$ 0.39\%), confirming that cross-domain distillation specifically enhances generalization and mitigates in-domain overfitting. 


\noindent\textbf{Movement-Guided Pruning.} 
This component proves critical in challenging scenarios. Although removing it yields minimal impact on well-solved benchmarks like 21DF (2.76\% $\rightarrow$ 3.13\%), its essential role becomes evident in more challenging scenarios, causing a noticeable degradation on ASVspoof5 (17.75\% $\rightarrow$ 19.95\%) and FoR (10.51\% $\rightarrow$ 13.56\%). We attribute this to the different difficulty levels across datasets. ASVspoof5 and FoR introduce more complex attack patterns~\cite{wang2024asvspoof5, reimao2019dataset-for} that are harder to distinguish. Particularly for FoR, its high-fidelity audio renders synthetic artifacts subtle for the pruned model to capture. This aligns with the core insights from prior work~\cite{hooker2019compressed}, which demonstrates that compression weakens the model's ability to handle high-difficulty samples, leading to a selective degradation on complex datasets. Our results confirm that movement-guided pruning could mitigate this forgetting behavior and preserve discriminative structures in challenging scenarios even under aggressive compression.
\begin{figure}[th]
  \centering
  \includegraphics[width=\linewidth]{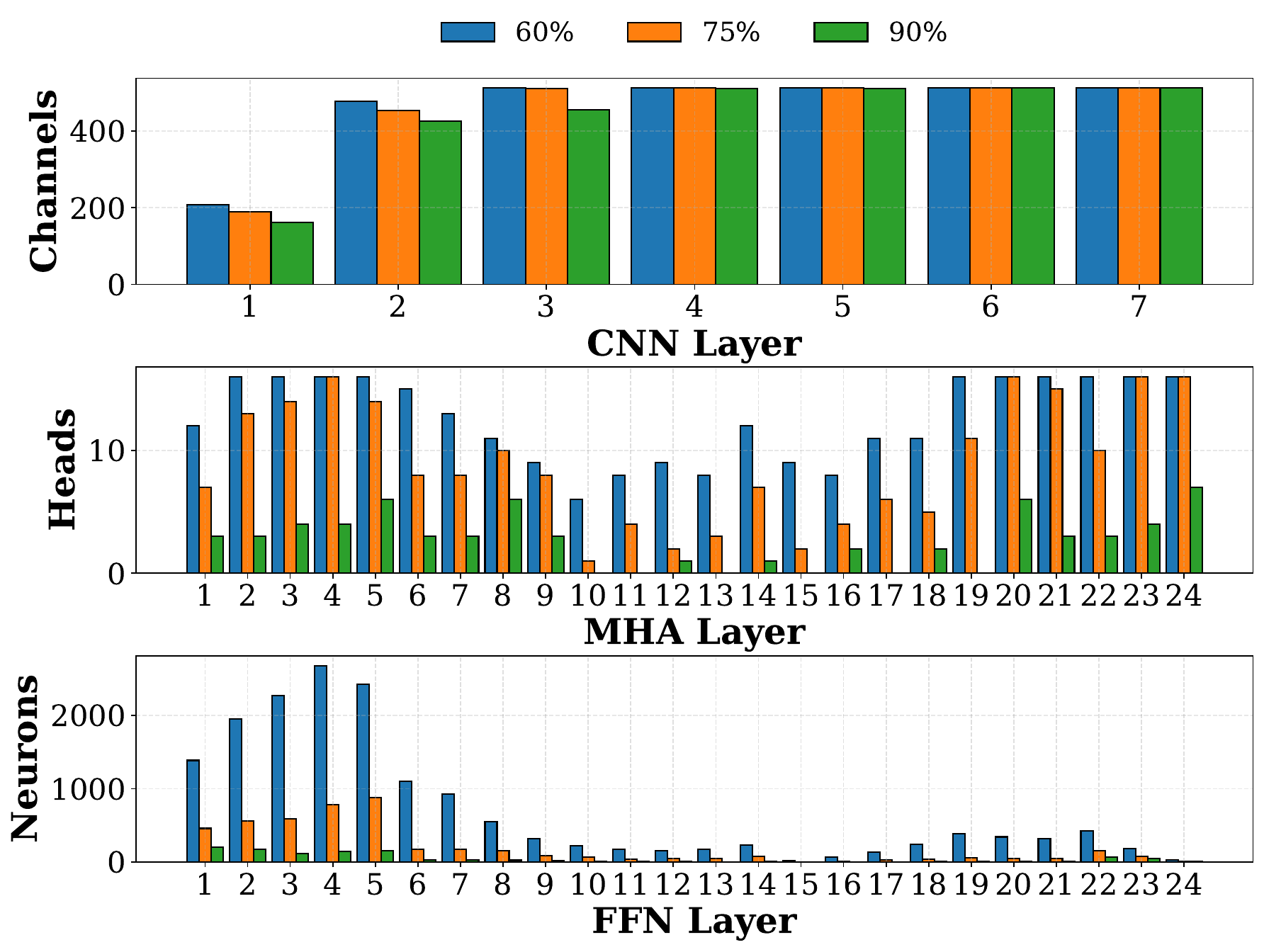}
  \vspace{-0.6cm} 
  \caption{The architecture of the pruned XLS-R at sparsity \{60\%,75\%,90\%\}. The original size of CNN channels, Attention heads, and FFN intermediates are 512, 16, and 4096.}
  \label{fig:visulize}
  \vspace{-0.5cm} 
\end{figure}

\subsection{Further Analysis}

\noindent\textbf{Architectural pruning patterns.} 
Figure~\ref{fig:visulize} reveals the pruned XLSR architecture. The CNN encoder retains most channels even at 90\% sparsity, confirming the foundational role of low-level spectral-temporal features for deepfake detection. MHA layers exhibit a ``U-shaped'' retention pattern, with layers 1--8 and 20--24 preserved, 9--19 compressed. FFN layers undergo extreme compression with intermediate dimensions shrinking to near-zero in deeper blocks, suggesting their parameterization is largely redundant for binary deepfake detection.

\noindent\textbf{Layer selection for distillation.} 
As shown in Figure~\ref{fig:cka_comparison}, CKA similarity analysis reveals block-like structures. Our additional experiments show that layer choices within the same block (e.g., $5$ vs $8$, $10$ vs $14$) yield comparable performance, suggesting that representative sampling from each stage matters more than exact indices. Future work could explore adaptive layer selection strategies, such as similarity-aware distillation \cite{zampierin2024skill}.


\noindent\textbf{Future directions.}
Our work suggests several promising directions for lightweight deepfake detectors.
First, although our framework is evaluated on XLS-R, it targets standard MHA and FFN blocks in the Transformer architecture, making it migratable to other SSL-based detectors. Future work could explore its cross-architecture portability and formulate more compression strategies for deepfake detection.
Second, our cross-domain distillation is inherently bounded by the teacher's errors on unlabeled data. We have attempted confidence-based calibration to mitigate this, but it yields marginal gains, necessitating further investigation into more robust distillation strategies. 
Third, as larger SSL models (e.g., XLS-R-2B~\cite{ge2025post}) achieve superior detection performance, scaling existing compression methods to such models remains an important open question.
\section{Conclusion}

We propose a Task-Aware Joint Pruning and Distillation framework for compressing SSL-based deepfake detection models. By combining cross-domain knowledge distillation with movement-guided structured pruning, our lightweight model preserves forgery-discriminative structure and maintains generalization across diverse scenarios under compression. Experiments demonstrate that we achieve 31.9M parameters and 6.3$\times$ FLOPs reduction while maintaining competitive performances, taking a promising step toward on-device deepfake detection.

\section{Acknowledgments}
This work was supported by the Shanghai Municipal Special Program for Basic Research on General AI Foundation Models (Grant No. 2025SHZDZX025G17), in collaboration with Shanghai Artificial intelligence Laboratory. It was also supported by the National Natural Science Foundation of China (No. 62472372 and No. 62572424)) and the Zhejiang Provincial Natural Science Foundation of China under Grant (No. LD24F020010 and No. LY24F020007).

\section{Generative AI Use Disclosure}
Generative AI tools were used solely for manuscript editing and polishing, with the authors retaining full responsibility and accountability for the entire work.

\bibliographystyle{IEEEtran}
\bibliography{bibs}

\end{document}